\documentclass[a4paper,manuscript]{aastex}
\usepackage[T1]{fontenc}
\usepackage[utf8]{inputenc}
\usepackage{float}
\usepackage{amssymb}
\usepackage{graphicx}

\makeatletter

\pdfpageheight\paperheight
\pdfpagewidth\paperwidth

\shorttitle{DBM fitting}
\shortauthors{Žic et al.}

\makeatother

\begin{document}

\sloppy

\global\long\def\d{\mathrm{d}}

\global\long\def\D{\mathrm{D}}

\global\long\def\vec#1{\mathbf{#1}}

\global\long\def\e{\mathrm{e}\,}

\global\long\def\sec{\mathrm{sec}\,}

\global\long\def\Me{\mathrm{Me}\,}

\title{Heliospheric Propagation of Coronal Mass Ejections:\\
 Drag-Based Model Fitting}

\author{T. Žic\altaffilmark{1} and B. Vršnak\altaffilmark{1}}

\email{\url{tzic@geof.hr}, \url{bvrsnak@geof.hr}}

\author{M. Temmer\altaffilmark{2}}

\email{\url{manuela.temmer@uni-graz.at}}

\altaffiltext{1}{Hvar Observatory, Faculty of Geodesy, University of Zagreb, Kačićeva
26, HR-10000 Zagreb, Croatia}

\altaffiltext{2}{Kanzelhöhe Observatory/IGAM, Institute of Physics, University of
Graz, Universitätsplatz 5, A-8010 Graz, Austria}

\begin{abstract}
The so-called drag-based model (DBM) simulates analytically the propagation
of coronal mass ejections (CMEs) in interplanetary space and allows
the prediction of their arrival times and impact speeds at any point
in the heliosphere (``target''). The DBM is based on the assumption
that beyond a distance of about 20 solar radii from the Sun, the dominant
force acting on CMEs is the ``aerodynamic'' drag force. In the standard
form of DBM, the user provisionally chooses values for the model input
parameters, by which the kinematics of the CME over the entire Sun--``target''
distance range is defined. The choice of model input parameters is
usually based on several previously undertaken statistical studies.
In other words, the model is used by ad hoc implementation of statistics-based
values of the input parameters, which are not necessarily appropriate
for the CME under study. Furthermore, such a procedure lacks quantitative
information on how well the simulation reproduces the coronagraphically
observed kinematics of the CME, and thus does not provide an estimate
of the reliability of the arrival prediction. In this paper we advance
the DBM by adopting it in a form that employs the CME observations
over a given distance range to evaluate the most suitable model input
parameters for a given CME by means of the least-squares fitting.
Furthermore, the new version of the model automatically responds to
any significant change of the conditions in the ambient medium (solar
wind speed, density, CME--CME interactions, etc.) by changing the
model input parameters according to changes in the CME kinematics.
The advanced DBM is shaped in a form that can be readily employed
in an operational system for real-time space-weather forecasting by
promptly adjusting to a successively expanding observational dataset,
thus providing a successively improving prediction of the CME arrival. 
\end{abstract}

\keywords{Sun: corona --- Sun: coronal mass ejections --- solar wind: solar-terrestrial
relations --- magnetohydrodynamics --- methods: analytical --- methods:
numerical}

\section{Introduction}

Eruptive processes in the solar atmosphere, particularly coronal mass
ejections (CMEs), strongly influence the physical state of the heliosphere
and the terrestrial space environment. CMEs represent eruptive restructuring
of the global coronal magnetic field, where the eruption itself is
caused by a loss of equilibrium of the pre-eruptive magnetic field
structure. The stability of the structure depends on the amount of
energy stored in the magnetic field, whereas the CME itself is driven
by the Lorentz force. The dynamics of the instability depends on the
magnetic-flux conservation and inductive effects, which cause the
cessation of the Lorentz force. Eventually, the magnetohydrodynamic
(MHD) drag becomes a dominant factor in the CME dynamics. The drag
is a consequence of collisionless transfer of momentum and energy
between the CME and the ambient solar wind by MHD waves \citep{Cargill2004_SoPh221_p135}.

In the present paper, we develop a method that provides the observations-driven
adjustment of the input parameters of the so-called drag-based model
(hereinafter, DBM), which describes the CME propagation in the interplanetary
space by considering the ``drag'' force (for details see \citet{Vrsnak2013_SoPh285_p295},
and references therein). The ``drag'' force depends on the relative
speed of the ejection and the solar wind; in a collisionless environment
the acceleration can be expressed as $a=-\gamma\left(v-w\right)\left|v-w\right|$,
where $\gamma$ is the ``drag parameter'', $a$ and $v$ refer to
the instantaneous acceleration and speed of the ejection, whereas
$w$ represents the ambient solar wind speed \citep{Vrsnak2001_SoPh202_p173,Cargill2004_SoPh221_p135,Owens2004_AnGeo22_p661,Vrsnak2007_AA472_p937,Borgazzi2009_AA498_p885,Lara2009_IAUS257_p287,Vrsnak2010_AA512_pA43,Vrsnak2013_SoPh285_p295}.
Furthermore, the previously used DBM with constant $\gamma$ and $w$
parameters \citep{Vrsnak2013_SoPh285_p295} is extended into a more
general form, allowing variable $\gamma(r)$ and $w(r)$. In the DBM
the CME is represented by the cone shape, where each element of the
CME's leading edge is defined by its position relative to the CME
tip. The parameters $\gamma$ and $w$ represent the most sensitive
elements of the DBM and play the main role in the drag-based simulation
of heliospheric CME propagation. Consequently, their evaluation represents
the central issue in DBM-based space-weather forecasting.

The paper is focused on the theoretical elaboration of finding values
of the DBM parameters that give the smallest difference between the
DBM-based kinematics and the CME kinematics as derived from observational
data. The observational measurements could be derived from coronagraphic
and heliospheric imaging data using several methods based on certain
assumptions (e.g., fixed $\varphi$ for small CMEs: see \citealt{Sheeley1999_JGR104_p24739,Rouillard2008_GeoRL35_p10110}
or harmonic mean for large CMEs: see \citealt{Lugaz2009_AnGeo27_p3479}).
The presented fitting method opens the possibility of an ``automatic''
evaluation of the most appropriate DBM input parameters from observational
data available for a particular event. The application and validation
of the proposed method will be presented in a follow-up paper employing
detailed coronal and heliospheric observations of one slow and one
fast CME.

\section{General description of the drag-based model}

\subsection{The drag force }

In interplanetary space the CME motion is governed by the Lorentz
force $F_{\mathrm{L}}$, gravity\textbf{ $F_{\mathrm{g}}$}, and the
MHD analog of the aerodynamic drag $F_{\mathrm{d}}$ \citep{Vrsnak2006_AdSpR38_p431}.
The net CME force can be expressed as: 
\begin{equation}
F=F_{\mathrm{L}}-F_{\mathrm{g}}+F_{\mathrm{d}}.\label{eq:F}
\end{equation}
At heliocentric distances beyond $R\gtrsim15$, the MHD drag becomes
a dominant force \citep{Vrsnak2009_IAUS257_p271}, so the CME motion
is basically influenced solely by the $F_{\mathrm{d}}$ term of the
force Equation~(\ref{eq:F}).

Generally, the ``drag'' interaction between the solar wind and the
CME in interplanetary space can be described in various ways. In this
paper we consider the ``drag'' force of the form: 
\begin{equation}
F_{\mathrm{d}}=-c_{\mathrm{d}}\,A\,\rho\,\left(v-w\right)\left|v-w\right|.\label{eq:Fd}
\end{equation}
where $c_{\mathrm{d}}$ refers to the dimensionless drag coefficient,
$A$ is the cross section of the CME, $\rho$ represents the ambient
solar wind density, and $(v-w)$ is the velocity difference between
the CME and the solar wind \citep{Chen1989_ApJ338_p453,Chen1993_GeoRL20_p2319,Cargill1996_JGR101_p4855,Cargill2000_JGR105_p7509,Cargill2002_AnGeo20_p879,Vrsnak2002_JGRA107_p1019,Vrsnak2004_AA423_p717,Vrsnak2009_IAUS257_p271,Vrsnak2010_AA512_pA43}.

Following the numerical MHD simulations by \citet{Cargill2004_SoPh221_p135},
and under the assumption that the CME structure does not change, we
expect that the drag coefficient $c_{\mathrm{d}}$ varies slowly with
radial distance and is approximately equal to 1 for the heliocentric
distances beyond 15 solar radii, particularly in the case of dense
CMEs. The mass density of a CME lies in the range of $12.7-13.5\,\mathrm{g}/r_{\odot}^{2}$,
with the most dense events occurring during the solar maximum (see
\citet{Vourlidas2011_ApJ730_p59}). In this respect, we define dense
CMEs as events with a mass density exceeding $13.2\,\mathrm{g}/r_{\odot}^{2}$.

The CME acceleration, caused by the MHD ``drag'' \citep{Vrsnak2009_IAUS257_p271},
can be written in a simple form using Equation~(\ref{eq:Fd}): 
\begin{equation}
a_{\mathrm{d}}=-\gamma\left(v-w\right)\left|v-w\right|,\label{eq:ad}
\end{equation}
where the parameter $\gamma$ is defined by 
\begin{equation}
\gamma=c_{\mathrm{d}}\frac{A\rho}{M}.\label{eq:gamma}
\end{equation}
The parameter $\gamma$ is inversely proportional to the total CME
mass $M$, which consists of the initial mass and the so-called virtual
mass that piles up as the CME expands in the inner heliosphere. Observations
indicate that beyond heliocentric distances of several solar radii
the total mass becomes approximately constant \citep{Bein2013_ApJ768_p31},
implying that the mass pile-up becomes balanced by the mass loss \citep{Vrsnak2007_AA472_p937,Vrsnak2013_SoPh285_p295}.

\subsection{Ambient density and solar wind speed}

For the ambient density $\rho_{0}(r)=m_{\mathrm{p}}\,n_{0}(r)$, where
$n_{0}(r)$ is unperturbed particle density and $m_{\mathrm{p}}$
is proton mass, the empirical $n_{0}(r)$ model proposed by \citet{Leblanc1998_SoPh183_p165}
is applied (referred to in the following as LDB, after Leblanc, Dulk,
Bougeret). The CME cross section $A(r)$ depends on the geometrical
shape of the CME; hereafter the cone representation is employed \citep{Fisher1984_ApJ280_p428,Xie2004_JGRA109_pA03109,Schwenn2005_AnGeo23_p1033}.
The parameter $\gamma(r)$ in Equation~(\ref{eq:ad}) defines the
effectiveness of the drag force and depends on both the CME and the
ambient solar wind.

At distances beyond $R\gtrsim15$ ($R\equiv r/r_{\odot}$, where $r_{\odot}$
is the solar radius), the terms $\propto R^{-4}$ and $\propto R^{-6}$
in the LDB expression for $n_{0}(R)$ can be neglected. However, for
the purposes of completeness and model development, in this paper
the complete LDB density expression is applied:

\begin{equation}
n_{0}(R)=\frac{k_{2}}{R^{2}}+\frac{k_{4}}{R^{4}}+\frac{k_{6}}{R^{6}},\label{eq:n0(R)}
\end{equation}
which is valid for radial distances $R>1.8$. Coefficients $k_{2}$,
$k_{4}$ and $k_{6}$ read $k_{2}=3.3\times10^{5}\,\mathrm{cm^{-3}}$,
$k_{4}=4.1\times10^{6}\,\mathrm{cm^{-3}}$ and $k_{6}=8.0\times10^{7}\,\mathrm{cm^{-3}}$
\citep{Leblanc1998_SoPh183_p165}.

\begin{figure}[h]
\noindent \begin{centering}
\includegraphics[width=1\columnwidth]{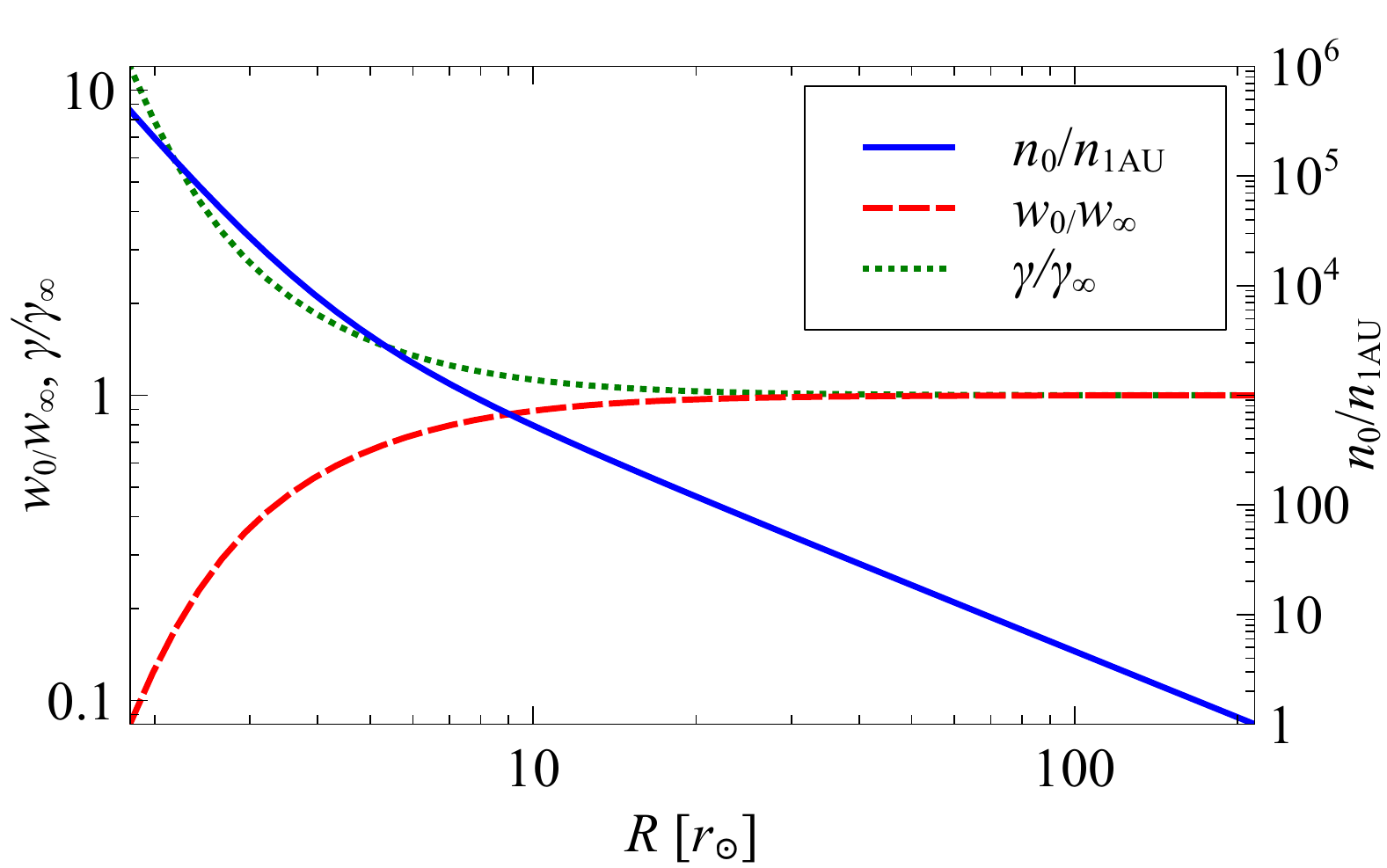}
\par\end{centering}

\noindent \centering{}\protect\caption{\label{fig:n-w-gamma}Radial dependence of the solar-wind density
$n_{0}$ normalized by the density $n_{\mathrm{1AU}}$ at 1~AU ($n_{0}/n_{\mathrm{1AU}}$;
solid line), shown together with the solar wind speed $w_{0}$ and
the drag parameter ratio $\gamma$ normalized by their asymptotic
values $w_{\infty}$ and $\gamma_{\infty}$ (dashed and dotted lines,
respectively).}
\end{figure}

The background solar wind is taken to be approximately stationary
and isotropic, so from flux conservation, $\partial n_{0}/\partial t+\nabla\cdot(n_{0}\vec w_{0})=\vec 0$,
the solar wind speed must satisfy the expression: 
\begin{equation}
w_{0}(R)=w_{\infty}\left(1+\frac{k_{4}/k_{2}}{R^{2}}+\frac{k_{6}/k_{2}}{R^{4}}\right)^{-1},\label{eq:w0(R)}
\end{equation}
where $w_{\infty}$ is the asymptotic solar wind speed, i.e., $w_{\infty}=\lim_{R\rightarrow\infty}w_{0}(R)=const$.
At small heliocentric distances the solar wind rarefies at a rate
larger than $R^{-2}$, so the wind speed has to rise according to
the continuity equation. Figure~\ref{fig:n-w-gamma} shows the radial
dependences of the normalized drag parameter $\gamma$, solar wind
speed $w_{0}$ and density $n_{0}$. The presented ratios $\gamma(R)/\gamma_{\infty}$
and $w_{0}(R)/w_{\infty}$ are normalized by asymptotic values ($\gamma_{\infty}$
and $w_{\infty}$), and the ratio $n_{0}(R)/n_{\mathrm{1AU}}$ by
the density value $n_{\mathrm{1AU}}$ at 1~AU. Note also that the
value of $\gamma$ close to the Sun is for an order of magnitude larger
than it is at large distances. As can be seen from Figure~\ref{fig:n-w-gamma},
one finds that the dependences $w_{0}(R)$, as well as $\gamma(R)$,
become practically constant beyond $R\gtrsim15$. Thus, the asymptotic
values of $w_{0}$ and $\gamma$ are approximately equal to the values
at 1~AU, i.e., $w_{\infty}\approx w_{\mathrm{1\,AU}}$ and $\gamma_{\infty}\approx\gamma_{\mathrm{1\,AU}}$.
A similar simplification was used in previous papers \citep{Vrsnak2007_AA472_p937,Vrsnak2010_AA512_pA43,Vrsnak2013_SoPh285_p295}
where the unperturbed solar wind speed and parameter-$\gamma$ functions
had constant values, $w_{0}(R)=w_{\infty}$ and $\gamma(R)=\gamma_{\infty}$,
for all radial distances.

The solar wind speed $w_{0}(R)$ can be additionally modified by including
a specific perturbation $w_{\mathrm{p}}(R)$ on top of the described
undisturbed background to reproduce a particular situation in a given
event. For example, in some cases the CME travels in interplanetary
space through a region of locally enhanced or decreased solar-wind
density \citep{Temmer2011_ApJ743_p101,Temmer2012_ApJ749_p57,Maricic2014_SoPh289_p351,Rollett2014_ApJL790_pL6}.
In such a case the additional $w_{\mathrm{p}}(R)$ term should describe
the associated solar wind speed perturbation in the region between
the heliocentric distances $R_{1}$ and $R_{2}$. Under these assumptions,
the perturbed solar wind speed is defined as 
\begin{equation}
w(R)=\left\{ \begin{array}{cc}
w_{0}(R)+w_{\mathrm{p}}(R), & R_{1}<R<R_{2}\\
w_{0}(R), & \mathrm{otherwise}
\end{array}\right.\label{eq:w(R)}
\end{equation}
where $w_{0}(R)$ represents the unperturbed solar wind speed (see
Equation~(\ref{eq:w0(R)})). The perturbed density induced by the
solar wind term $w_{\mathrm{p}}$ in $w(R)$, 

\begin{equation}
n(R)=\frac{k_{2}}{R^{2}}\frac{w_{\infty}}{w(R)},\label{eq:n(R)}
\end{equation}
follows from flux conservation, i.e., $n(R)=\lim_{R_{\infty}\rightarrow\infty}[n_{0}(R_{\infty})\,R_{\infty}^{2}\,w_{0}(R_{\infty})/w(R)\,R^{2}]$.
The perturbation is assumed to be localized over a finite region (i.e.,
inside the interval $R_{1}<R<R_{2}$), whereas the unperturbed expressions
for the density, Equation~(\ref{eq:n0(R)}), and solar wind speed,
Equation~(\ref{eq:w0(R)}), are valid otherwise.

The solar wind perturbation above is described by defining the wind
speed $w_{\mathrm{p}}(R)$ from which follows the density profile
$n(R)$; however, if the case study requires, the perturbation could
be performed in the opposite way, firstly defining the density perturbation
$n_{\mathrm{p}}(R)$ and afterward evaluating the solar wind speed
expression $w(R)$.

\subsection{Drag parameter $\gamma$}

In the case of the CME propagation where the ``drag'' force is dominant,
the equation of motion, Equation~(\ref{eq:ad}), transforms to

\begin{equation}
\ddot{R}(t)=-\gamma(R)\left[\dot{R}(t)-w(R)\right]\left|\dot{R}(t)-w(R)\right|.\label{eq:DE_a(T)}
\end{equation}

The expression for $\gamma$ given by Equation~(\ref{eq:gamma})
shows that the drag is more effective if the ambient density is high,
if the CME is light, and if the CME cross section is large. In the
present version of the DBM the effective CME cross section is defined
by employing the CME ``cone model'' \citep{Fisher1984_ApJ280_p428,Xie2004_JGRA109_pA03109,Schwenn2005_AnGeo23_p1033}.
In this presentation, the cross-sectional area is given by $A=\pi\,r_{\odot}^{2}\,R^{2}\,\tan^{2}\lambda/(1+\tan\lambda)^{2}$,
where $\lambda$ is the CME half-width (see the Appendix). Note that
CMEs could be represented by a variety of geometrical representations;
for examples of commonly used geometries see \citet{Schwenn2005_AnGeo23_p1033},
\citet{Thernisien2006_ApJ652_p763}, \citet{Lugaz2010_ApJ715_p493},
\citet{Thernisien2011_ApJS194_p33}, \citet{Davies2012_ApJ750_p23},
and the references therein.\textbf{ }Note that in the previous form
of DBM with constant $w$ and $\gamma$ \citep{Vrsnak2007_AA472_p937,Vrsnak2010_AA512_pA43,Vrsnak2013_SoPh285_p295}
the area was approximated by $A\approx\pi\,r_{\odot}^{2}\,R^{2}\,\lambda^{2}$.
However, this did not have a direct influence on the calculated CME
kinematics, since the value of $A$ was already incorporated within
the presumed parameter $\gamma$.

Taking into account the definition of solar wind speed, Equation~(\ref{eq:w(R)}),
$\gamma$ includes even the cases of perturbed solar wind, i.e., when
the term $w_{\mathrm{p}}(R)$ is taken into account: 
\begin{equation}
\gamma(R)=\gamma{}_{\infty}\frac{w_{\infty}}{w(R)}.\label{eq:gamma(R)}
\end{equation}
Since the asymptotic value of the solar wind speed at large heliocentric
distances ($R\rightarrow\infty$) is $w_{\infty}$ (see Figure~\ref{fig:n-w-gamma}),
evidently $\gamma(R)$ asymptotically acquires value $\gamma_{\infty}$
likewise,\emph{ }i.e., $\gamma_{\infty}=\lim_{R\rightarrow\infty}\gamma(R)$.
In space-weather forecasting it became a practice to use a dimensionless
variant $\Gamma$, defined by $\gamma_{\infty}=\Gamma\times10^{-7}\,\mathrm{km^{-1}}$. 

An interesting consequence follows from the dependence of the parameter
$\Gamma$ on the CME's geometrical shape and its properties. In the
case when the effective cross section $A$ is proportional to the
distance squared, $A\propto R^{2}$, $\Gamma$ could be generally
calculated using $\Gamma=f(\lambda)/M$. The expression $f(\lambda)$
depends on the CME geometrical shape of the CME used in the model,
which in our case is given by $f(\lambda)=\pi m_{p}r_{\odot}^{2}\,k_{2}\,\left[\tan\lambda/\left(1+\tan\lambda\right)\right]^{2}\times10^{7}\,\mathrm{km}=8.4\times10^{12}\,\left[\tan\lambda/\left(1+\tan\lambda\right)\right]^{2}\,\mathrm{kg}$
(see the Appendix). If the observations provide the CME half-width
angle $\lambda$, and observed kinematics provide the value of $\Gamma$,
the CME mass $M$ can be roughly estimated by employing $M=8.4\times10^{12}\,\left[\tan\lambda/\left(1+\tan\lambda\right)\right]^{2}\,\Gamma^{-1}\,\mathrm{kg}$,
where $\lambda$ is expressed in radians and $M$ in kg. The same
holds for the opposite situation: knowing the mass $M$ one can estimate
the angular half-width $\lambda$ from the value of $\Gamma$. The
presumed geometrical shape of the CME affects the estimation of cross-sectional
area, and consequently is important in evaluation of the unknown properties
($M$ or $\lambda$) of the CME. Thus, the best suited choice of the
geometrical model to the CME observational properties improves the
accuracy of the CME mass or half-width evaluation.

\section{Model/observations fitting}

In the following, the procedure of finding the values of any unknown
DBM parameters is described. The drag parameter $\Gamma$, the background
solar wind speed $w_{\infty}$, and the modified initial CME radial
distance $R_{0}$ and speed $v_{0}$ are adjusted iteratively by minimizing
the deviation of the model kinematics from the observed one. The process
sequentially alters the DBM parameters in order to minimize the quadratic
deviation (the sum of squared ``errors'' or residuals) between observational
and DBM-calculated speeds: 
\begin{equation}
E(\Gamma,w_{\infty};R_{0},v_{0})=\sum_{i=0}^{N}\left[v_{i}-v(\{\Gamma,w_{\infty};R_{0},v_{0}\},R_{i})\right]^{2}.\label{eq:E}
\end{equation}
The observational distance--speed values are written as $R_{i}$ and
$v_{i}$, while the adjusted kinematic curve $v(R)$, dependent on
the model input parameters $\Gamma$, $w_{\infty}$ and the initial-state
parameters $R_{0}$, $v_{0}$, is designated as $v(\{\Gamma,w_{\infty};R_{0},v_{0}\},R)$.
The initial-state parameters depend on the presumed geometrical representation
of a CME; therefore for the geometrical option presented in the Appendix,
the initial-state parameters $R_{0}$ and $v_{0}$ of Equation~(\ref{eq:E})
can have the CME tip values $R_{0}(t_{0})$, $v_{0}(t_{0})$, or the
flank values $R_{\varphi}(t_{0})$, $v_{\varphi}(t_{0})$ at the initial
time $t_{0}$, depending on the observer's location (see the Appendix).
Unknown input parameters are found by successively solving the equation
of motion, Equation~(\ref{eq:DE_a(T)}), within the parameter domain.
In practice we use a more appropriate form of Equation~(\ref{eq:DE_a(T)}),
which reads 
\begin{equation}
v(R)\,\frac{\d v(R)}{\d R}=-\gamma(R)\left[v(R)-w(R)\right]\left|v(R)-w(R)\right|.\label{eq:DE_v(R)}
\end{equation}
The variation of the DBM parameters seeks the minimal value $E_{\mathrm{min}}$
of Equation~(\ref{eq:E}). The presented method is basically a modified
successive multiparametric variation that includes solving of the
differential equation of motion, Equation~(\ref{eq:DE_v(R)}), and
least-squares fitting (hereafter, LSF) to the observational $\{(R_{0},v_{0}),\ldots,(R_{N},v_{N})\}$
dataset. Different approaches could be used in the numerical fitting.
For example, the computation could be performed by starting with arbitrary
DBM values based on which optimal values are found, or by numerically
seeking the minimum of Equation~(\ref{eq:E}) within a physically
meaningful DBM-parameter domain \citep{Motulsky1987_FASEBJ1_p365}.
The meaningful parameter-domain restriction could be also included
in the firstly mentioned approach to speed up the process of finding
the $E_{\mathrm{min}}$. In the end, the minimal quadratic deviation
gives the best input-parameter set $\{\Gamma,w_{\infty};R_{0},v_{0}\}$
for the specific observational event. Furthermore, kinematic curves,
such as $a(R)$, $a(t)$, $v(R)$, $v(t)$, and $R(t)$ are automatically
available from the calculated parameters. Consequently, this directly
provides the CME transit time $\tau$, defined as the time the CME
takes to arrive at a prescribed location, as well as the ``impact''
velocity $v_{\tau}$.

It is instructive to employ statistical analysis and to express the
``goodness'' of fit in the form of several statistical quantities.
The first is\emph{ }the standard deviation\textbf{ }(or the rms),
which represents the average deviation between observed \textbf{$v_{i}$}
and calculated $v(R_{i})$ data:

\begin{equation}
\sigma=\sqrt{\frac{\sum_{i=0}^{N}\left[v_{i}-v(R_{i})\right]^{2}}{N+1}},
\end{equation}
and gives data dispersion in velocity units (i.e. $\mathrm{km\,s^{-1}}$).
Notice that the $(N+1)$ is the total number of observational datapoint
samples $\{(R_{0},v_{0}),\ldots,(R_{N},v_{N})\}$. 

The ``goodness'' could be graphically presented in the form of a
residual plot. A residual plot shows the differences (or residuals)
between each measured $v_{i}$ value and the value calculated from
the estimated curve $v(R_{i})$, i.e., $[v_{i}-v(R_{i})]$. The residuals
should not have a systematic dependence on $R$ values (the abscissa
values) and should have a random scattering. Any clustering of residuals
in the plot indicates that the prediction curve follows a systematic-error
pattern and that the fit is not appropriate.

The next relevant criterion of scattering between observed values
$v_{i}(R_{i})$ and calculated $v(R)$ is the coefficient of variation,
which is defined by
\begin{equation}
c_{v}=\frac{\sigma}{\bar{v}}\cdot100\%,
\end{equation}
where $\bar{v}=\sum_{i=0}^{N}v(R_{i})/(N+1)$ is the mean of calculated
values from the calculated set $\{v(R_{i})\}$.

Lastly, the coefficient of determination is defined by
\begin{equation}
\mathcal{R}^{2}=1-\frac{\sum_{i=0}^{N}\left[v_{i}-v(R_{i})\right]^{2}}{\sum_{i=0}^{N}\left[v_{i}-\bar{v}\right]^{2}},\quad0\le\mathcal{R}^{2}\le1
\end{equation}

The ``quality'' of the estimated DBM parameters increases, i.e.,
the calculated kinematic curve fits observational data better, as
$\sigma$ and $c_{\mathrm{v}}$ decrease and reach minimal values
$\sigma_{\mathrm{min}}$ and $c_{\mathrm{vmin}}$, respectively. On
the other hand, $\mathcal{R}^{2}$ becomes close to 1 as fit gets
better \citep{Motulsky1987_FASEBJ1_p365}.

\begin{figure}[H]
\begin{centering}
\includegraphics[width=1\columnwidth]{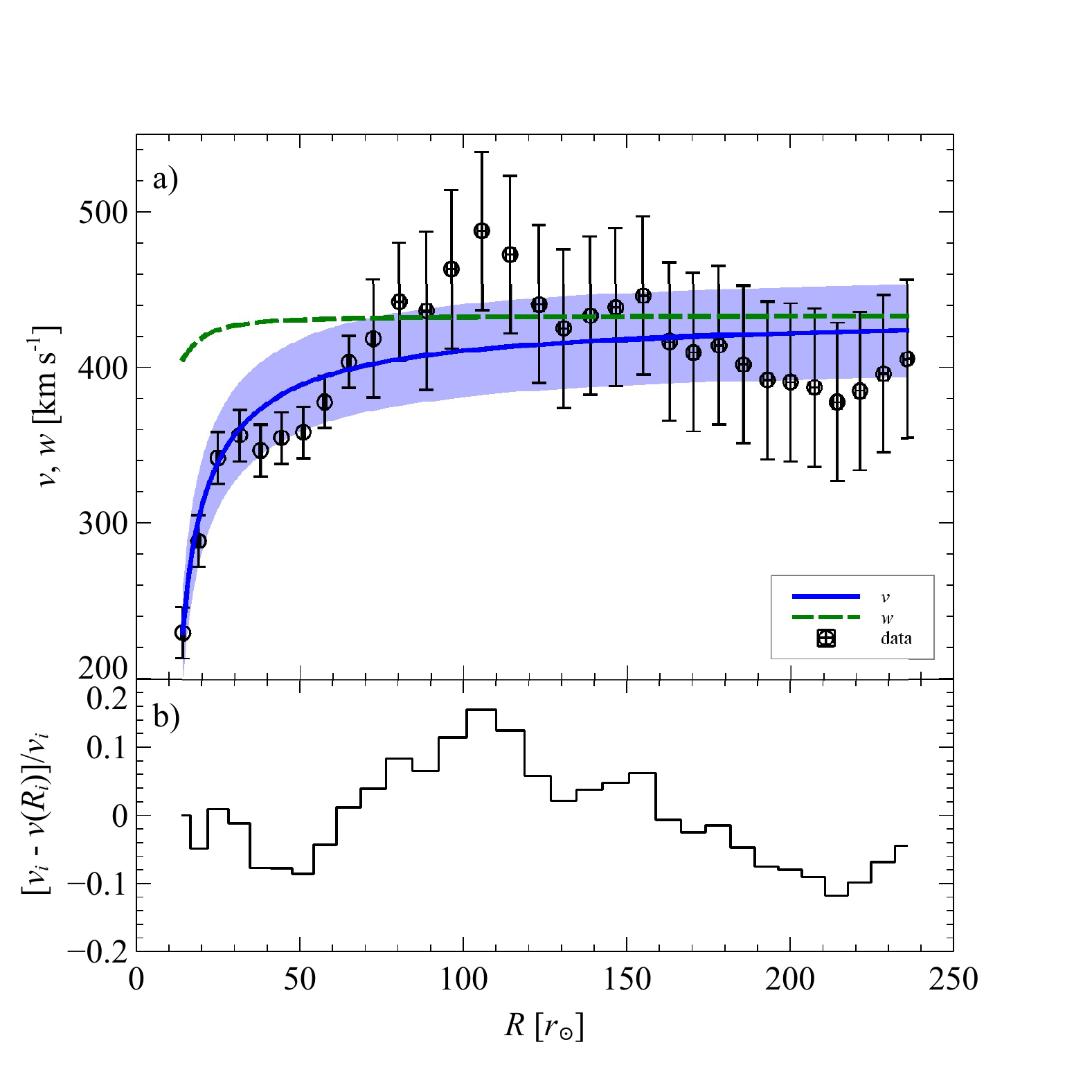}
\par\end{centering}

\protect\caption{\label{fig:LSF-DBM demo}Application of the LSF--DBM method on the
dataset extracted from Event~1 of the paper by \citet{Temmer2011_ApJ743_p101}.
(a)~Radial dependence of the LSF--DBM estimated kinematic curve $v(R)$
(solid blue line with error shown as the blue shaded area) on the
observed CME speed values (black circles, with error bars), and numerically
calculated solar wind speed $w$ (dashed green) based on parameter
fitting. (b)~Radial dependance of residuals, $v_{i}-v(R_{i})$, between
observational $v_{i}$ and DBM-calculated $v(R_{i})$ speeds of CME
(relative to observed CME speed $v_{i}$).}
\end{figure}

For demonstration purposes we have chosen the observational dataset
of Event~1 described in \citet{Temmer2011_ApJ743_p101} and we have
applied the LSF--DBM method to simulate the CME propagation. The event
started on 2008 June 1 at $\sim21\,\mathrm{UT}$ and the propagation
data $\{R_{i},v_{i}\}$ with associated errors are derived from \emph{STEREO}
coronagraphic and heliospheric image data using the constrained harmonic
mean method (see \citealt{Rollett2012_SoPh276_p293}). For more details
we refer to \citet{Temmer2011_ApJ743_p101}. The final result of the
DBM fitting is presented in Figure~\ref{fig:LSF-DBM demo}, where
panel \ref{fig:LSF-DBM demo}(a) presents the velocity--distance profile
of the calculated CME kinematics (blue solid line and accompanying
shaded error area), as well as the estimated solar wind speed (green
dashed line), together with the observational dataset (black circles)
and its error bars. In the bottom panel (\ref{fig:LSF-DBM demo}(b))
we present the residulas, i.e.\emph{,} the relative difference $[v_{i}-v(R_{i})]/v_{i}$
between observational, $v_{i}$, and calculated, $v(R_{i})$, CME
velocities, relative to the velocity $v_{i}$. The standard deviation
of the observed dataset is $\sigma_{\mathrm{o}}=\sqrt{\sum_{i=0}^{N}e_{i}^{2}/(N+1)}=42.81\,\mathrm{km\,s^{-1}}$,
where $e_{i}$ represents the half-error bar value for each measurement
of velocity $v_{i}$. The LSF--DBM technique produced the fit with
the DBM parameters $\Gamma=2.84$, $w_{\infty}=433.04\,\mathrm{km\,s^{-1}}$,
$v_{0}=229.50\,\mathrm{km\,s^{-1}}$, $R_{0}=14.17\,r_{\odot}$, accompanied
by the minimal standard deviation and the coefficients of variation
and determination, $\sigma_{\mathrm{min}}=29.87\,\mathrm{km\,s^{-1}}$,
$c_{\mathrm{vmin}}=7.50\,\%$, $\mathcal{R}^{2}=0.67$, respectively.
In Figure~\ref{fig:LSF-DBM demo}(a) the ``average'' errors of
the fit, spanning values frim $v(R)-\sigma_{\mathrm{min}}$ to $v(R)+\sigma_{\mathrm{min}}$,
are drawn as the blue shaded area in the vicinity of the DBM kinematic
curve $v(R)$. Evidently, the fitted standard deviation $\sigma_{\mathrm{min}}$
is much smaller than the observed $\sigma_{\mathrm{o}}$, showing
that the LSF--DBM produced a satisfactory fit within the range of
the ``average observational error'' $\sigma_{\mathrm{o}}$.

\section{Discussion and conclusion}

We have presented an extension of the DBM that is intended for use
in automatic forecasting of CME arrival and impact at an arbitrary
heliospheric position. The extension consists of optimizing the DBM
input parameters based on the sequential variation and determination
of the minimal standard deviation ($\sigma_{\mathrm{min}}$), the
minimal coefficient of variation ($c_{\mathrm{vmin}}$), and the coefficient
of determination ($\mathcal{R}^{2}$) from observational data. The
mentioned statistical quantities represent an estimate of ``goodness''
of the DBM fit to observational data and consequently define the reliability
of the arrival time and impact speed prediction. The presented LSF--DBM
modification opens an opportunity for implementation in real-time
space-weather forecasting tools and alerting systems for CME impacts
on Earth (or any heliospheric ``target'' of interest). The novel
approach is based on real-time data-driven DBM-parameter optimization
that iteratively improves the accuracy of CME kinematics in the heliosphere.

The accuracy of the real-time forecasting increases as the observational
dataset successively becomes larger, i.e., as the CME is tracked to
larger heliospheric distances \citep{Davis2010_SoPh263_p209}. For
example, we can imagine a hypothetical case study of a CME launched
from a region close to the solar disc center at a specific time $t=0$.
In this case the CME is directed toward Earth and we can extract information
about a current in situ solar wind speed $w_{\infty}$ at the Earth.
On the other hand, the CME is traced in real time during its propagation
throughout the heliosphere, so by using the most likely CME geometry
the observational data can be transformed to get the distance--speed
$(R_{i},v_{i})$ data. Every time when a new $(R_{i},v_{i})$ datapoint
become available, the LSF procedure estimates a new set of DBM parameters
required for updating the DBM forecast of CME arrival. As the dataset
expands, our ``impact prediction'' becomes more reliable. In this
respect it should be noted that an L5 mission is urgently needed to
advance the performance of such forecasting methods. We note that
this example is a quite simplified case in which the LSF--DBM method
could be used.

There are several drawbacks of the described procedure, e.g., the
estimation of the model input parameter and the related forecasting
are highly dependent on the quality of the observational dataset.
The input required for the DBM fitting procedure is the observed set
of values for speed and distance of the CME frontal part as observed
along the ecliptic plane. Several methods exist to derive those quantities,
and we just mention briefly some possibilities. The propagation direction
might be simply estimated from the CME associated source region, assuming
radial propagation. In fact, knowing the propagation direction would
enable one to derive the 3D CME kinematics from single spacecraft
observations, such as, e.g., from \emph{STEREO} heliospheric image
data assuming a certain CME width (e.g., fixed $\phi$ for small CMEs:
see \citealt{Sheeley1999_JGR104_p24739,Rouillard2008_GeoRL35_p10110},
or harmonic mean for large CMEs: see \citealt{Lugaz2009_AnGeo27_p3479}).
Using stereoscopic data, triangulation methods could be used that
also provide the required input (coronagraphic field of view: see
\citealt{Mierla2010_AnGeo28_p203}; interplanetary space: see \citealt{Liu2010_ApJL710_p82}).
The uncertainty of measurements is automatically forwarded to the
estimated model parameters, and consequently, to the arrival time
prediction.

Another serious drawback lies in the fact that the employed observational
data include the distance range where the CME is still driven by the
Lorentz force \citep{Gallagher2003_ApJL588_p53}. In such a situation,
the DBM fails in its fundamental concept because the Lorentz force
is excluded from the modeling, i.e., only the drag force governs the
CME propagation. However, it should be noted that even in such a case,
the DBM kinematical curve might fit the observational data nicely
due to the fact that the statistical weight dominantly comes from
larger heliospheric distances, where the Lorentz force should be negligible.
For example, if the observational dataset used in the modeling consists
of only a few low-height measurements and a more abundant subset of
measurements at larger heliocentric distances, the larger drag-dominated
dataset ``overweights'' the smaller Lorentz-driven subset, so the
latter effect becomes negligible.

The LSF and the DBM could be used in an opposite way to that previously
discussed, for example to estimate the solar wind speed $w_{\infty}$
at large heliospheric distances ($R\gtrsim15\,r_{\odot}$), in the
regions where in situ measurements are not available. Moreover, measuring
$(R_{i},v_{i})$ LSF straightforwardly gives the solar wind speed,
$w_{\infty}$. Using Equations~(\ref{eq:w(R)}) and (\ref{eq:n(R)})
we can then roughly calculate $w(R)$ and $n(R)$ for any heliocentric
distance, $R$. Additionally, as the LSF estimates the complete set
of DBM parameters, $\{\Gamma,w_{\infty};R_{0},v_{0}\}$, in situations
when the measurements are not very confident and have a high uncertainty,
we could apply the LSF method and correct, for example, the low-coronal
initial position and the velocity of a CME. However, the unknown DBM
parameters are more reliably estimated as more parameters are directly
given from the observations, and if stereoscopic observations are
conducted in an appropriate manner to provide reliable deprojected
($R_{i}$, $v_{i}$) values.

The LSF--DBM could be further applied in a case when a CME meets various
heliospheric ``obstacles'' during its propagation. The probability
of an interaction between two consecutive CMEs is very high in the
heliosphere, since on average several CMEs are observed per day with
different kinematics and velocities \citep{StCyr2000_JGR105_p18169,Gopalswamy2006_JApA27_p243}.
The interaction takes place when the later and faster CME catches
an earlier and slower one \citep{Temmer2012_ApJ749_p57,Maricic2014_SoPh289_p351}.
By inspecting the CME's behavior and surrounding ambient conditions,
the LSF--DBM procedure could be used for a ``segmented-distance''
application. For example, the CME trajectory could be divided into
several parts dependent on the CME behavior, i.e., divided into regions
before the CME--CME interaction and the region after the interaction.
In that way the forecasting of the CME arrival at a given ``target''
could be acquired by applying sequentially the LSF--DBM technique
on each trajectory interval (e.g., \citealt{Temmer2011_ApJ743_p101,Temmer2012_ApJ749_p57,Maricic2014_SoPh289_p351,Rollett2014_ApJL790_pL6}).

Actually, numerical computation requires arbitrary initial DBM-parameter
entries around which the LSF procedure searches for the best result.
Sometimes the problem arises when, in the proximity of starting DBM
entries, the numerical LSF finds multiple $\sigma_{\mathrm{min}}$
minima inside the parameter domain. The problem could be avoided by
carefully studying the specific case, or using a different track-fitting
method (see, e.g., \citealt{Mostl2013_SoPh_p285} who use a constant-velocity
approximation) and then reapply the LSF--DBM procedure to refine the
forecasting.

The presented generalized DBM is an extension of the model with the
assumption of a constant $\gamma(R)$ and $w(R)$ \citep{Vrsnak2013_SoPh285_p295},
which is not adequate for describing low-coronal CME propagation,
or kinematics in the spatially perturbed solar wind $w(R)$. Finally,
the application of the least-squares fitting method coupled with the
DBM applied to various CME geometries and solar wind models offers
an improvement in efficiency and accuracy of forecasting CME.

\acknowledgements{This work has been supported in part by the Croatian Science Foundation
under the project 6212 “Solar and Stellar Variability” and by the
European Union Seventh Framework Programme (FP7/2007-2013) under grant
agreements no. 263252 (COMESEP; \url{www.comesep.eu}) and no. 284461
(eHEROES; \url{http://soteria-space.eu/eheroes/html/}). M.~T. gratefully
acknowledges the Austrian Science Fund (FWF): FWF V195-N16. }

\appendix

\section{\label{sec:CME-geometry}CME shape used for DBM online tool}

\begin{figure}[H]
\begin{centering}
\includegraphics[width=0.75\columnwidth]{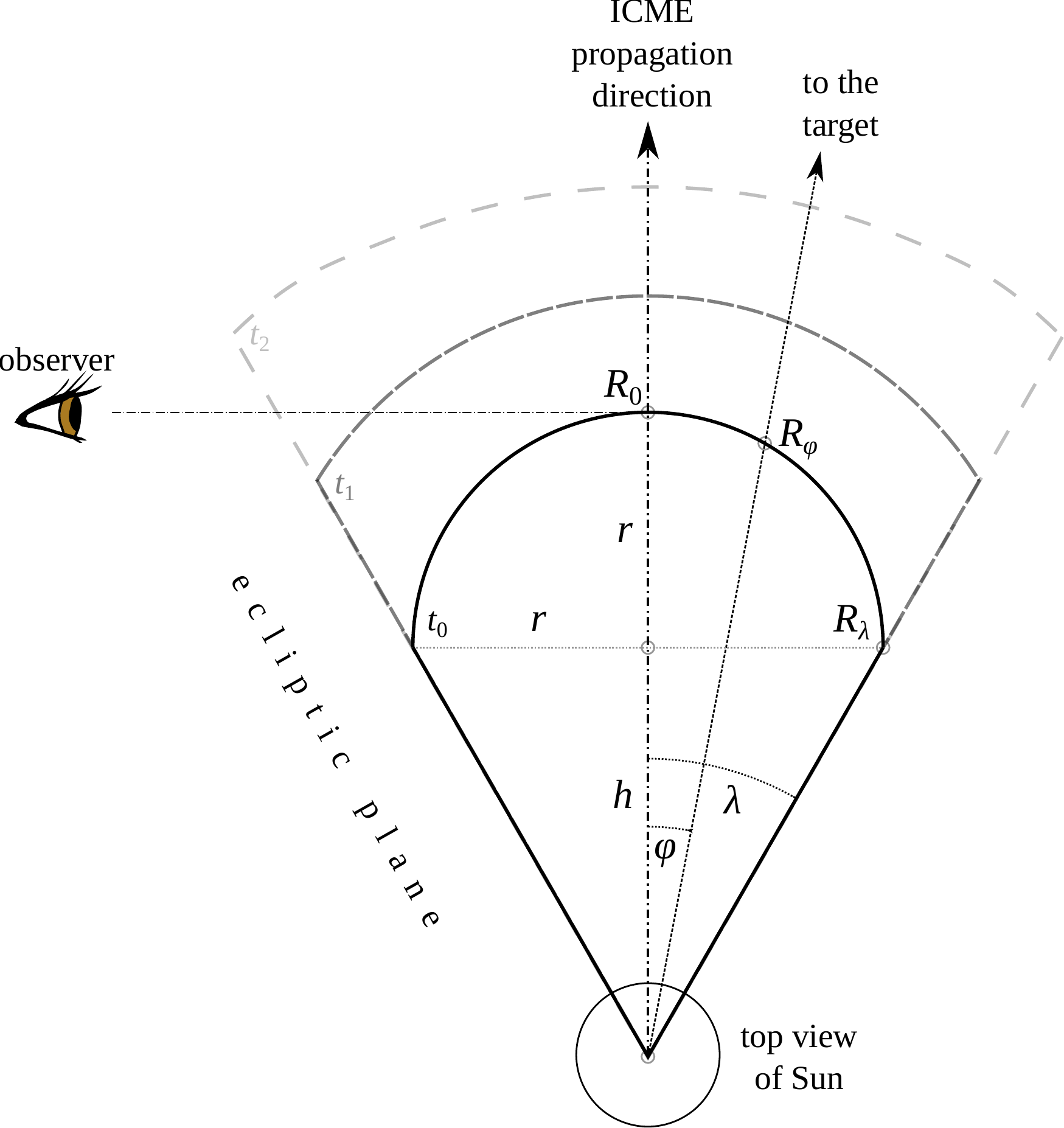}
\par\end{centering}

\protect\caption{\label{fig:Cone-model} A cross section of a conical representation
of a CME in the ecliptic plane during ICME propagation. The simple
schematic describes temporal deformation and evolution of the CME's
leading edge in time. The \emph{initial} CME shape at time $t_{0}$
is defined from a single heliocentric distance measurement $R_{0}$
of the CME tip and used in an equation that defines the conical geometry.}
\end{figure}

We briefly discuss the general outcome of a DBM calculation and its
dependence on the presumed shape of the leading edge of the interplanetary
CME (ICME). To evaluate the CME cross-sectional area, $A$, (and therefore
$\Gamma$) in this version of the DBM we used the geometry presented
in Figure~\ref{fig:Cone-model} (for other frequently used options
see, e.g., \citealp{Sheeley1999_JGR104_p24739,Kahler2007_JGRA112_p9103,Lugaz2009_AnGeo27_p3479,Davies2012_ApJ750_p23}
and the references therein). In this option the\emph{ }leading edge
is considered to be a semicircle, spanning over the full angular width
of the ICME, $2\lambda$. Considering the geometrical relationships
between various parameters marked in Figure~\ref{fig:Cone-model},
the\emph{ }heliocentric distance $R_{\varphi}(t)$ and the speed $v_{\varphi}(t)$
of an element at the angular position $\varphi$ depend on the heliocentric
distance of the CME tip, $R_{0}(t)$, the speed of the CME tip $v_{0}(t)$,
the cone half-width $\lambda$ (which stays constant during ICME propagation),
and the angle $\varphi$. Precisely, the relationships between radial
distances and velocities of the tip and a flank CME element are 
\begin{eqnarray}
R_{\varphi}(t) & = & R_{0}(t)\,F(\varphi)\nonumber \\
v_{\varphi}(t) & = & v_{0}(t)\,F(\varphi)\label{eq:R_phi=000026v_phi}
\end{eqnarray}
 respectively, where the angular function is the same for both expressions:
\begin{equation}
F(\varphi)=\frac{\cos\varphi+\sqrt{\tan^{2}\lambda-\sin^{2}\varphi}}{1+\tan\lambda}.
\end{equation}
The CME expansion is modeled by providing the initial speed $v_{0}$
and heliocentric distance $R_{0}$ of an arbitrary single point on
the CME's leading edge (e.g., in Figure~\ref{fig:Cone-model} the
leading edge segment of the CME tip has the distance $R_{0}(0)$ and
the speed $v_{0}(0)$), thus the heliocentric distances $R_{\varphi}(0)$
and speeds $v_{\varphi}(0)$ of a certain segment along the leading
edge with $\varphi\in[-\lambda,\lambda]$, follow from Equation~(\ref{eq:R_phi=000026v_phi}).
At later time the leading edge evolves accordingly, as described in
Equation~(\ref{eq:DE_a(T)}).

In the current versions of the DBM, implemented as a public prognostic
online tool at \url{http://www.geof.unizg.hr/~tzic/dbm.html}, the
different options of CME expansions are proposed. The prognostic tool
forecasts only ICME propagation in the ecliptic plane, as a result
of CME initiation at low heliographic latitudes. The geometric setup
of the latest online DBM version is presented in Figure~\ref{fig:Cone-model}
where the CME frontal part evolves in time, i.e., the expansion of
the CME's leading edge is simulated by applying the DBM equation of
motion, Equation~(\ref{eq:DE_a(T)}), on each leading-edge segment
independently. The initial cross section in the ecliptic plane of
the CME shape is constructed by a single $R_{0}(0)$ measurement of
the CME tip element (which lies on the line of the CME's propagation
direction and in the ecliptic plane as well) and by the assumption
of conical CME geometry defined by Equation~(\ref{eq:R_phi=000026v_phi}).
The leading edge gradually deforms since different segments are immersed
at initial time $t_{0}$ in different surrounding conditions (described
by solar wind speed $w(R)$ and $\gamma(R)$ functions) and have different
initial velocities (see Equation~(\ref{eq:R_phi=000026v_phi})),
hence the DBM equation of motion results in different radial kinematics.
Since the flanks move more slowly, and thus in fast ICMEs the drag-deceleration
of flanks is weaker whereas flank acceleration in slow events is stronger,
the variation of speed along the ICME front decreases and the front
gradually flattens. Note that such an ``independent-element'' DBM
procedure could be equivalently applied to any other presumed initial
CME geometry.

\bibliographystyle{aa}
\bibliography{biblio-abbr-ApJ,biblio}

\clearpage{}

\end{document}